\documentclass[twocolumn,preprintnumbers,pra]{revtex4-2}

\usepackage{graphicx}
\usepackage{amsfonts}
\usepackage{amsmath}
\usepackage[subrefformat=parens]{subcaption}
\usepackage{dcolumn}
\usepackage{bm}
\usepackage{braket}
\usepackage[T1]{fontenc} 

\begin{document}
\title{Resolution enhancement of one-dimensional molecular wavefunctions in plane-wave basis
via quantum machine learning}
\author{Rei Sakuma}
\email[]{rei\_sakuma@jsr.co.jp}
\affiliation{Materials Informatics Initiative, RD Technology \& Digital Transformation Center,
    JSR Corporation, 3-103-9 Tonomachi, Kawasaki-ku, Kawasaki, Kanagawa 210-0821, Japan}
\author{Yutaro Iiyama}
\email[]{iiyama@icepp.s.u-tokyo.ac.jp}
\affiliation{International Center for Elementary Particle Physics (ICEPP), The University of Tokyo, 7-3-1 Hongo, Bunkyo-ku, Tokyo 113-0033, Japan}
\author{Lento Nagano}
\email[]{lento@icepp.s.u-tokyo.ac.jp}
\affiliation{International Center for Elementary Particle Physics (ICEPP), The University of Tokyo, 7-3-1 Hongo, Bunkyo-ku, Tokyo 113-0033, Japan}
\author{Ryu Sawada}
\email[]{sawada@icepp.s.u-tokyo.ac.jp}
\affiliation{International Center for Elementary Particle Physics (ICEPP), The University of Tokyo, 7-3-1 Hongo, Bunkyo-ku, Tokyo 113-0033, Japan}
\author{Koji Terashi}
\email[]{terashi@icepp.s.u-tokyo.ac.jp}
\affiliation{International Center for Elementary Particle Physics (ICEPP), The University of Tokyo, 7-3-1 Hongo, Bunkyo-ku, Tokyo 113-0033, Japan}

\date{\today }

\begin{abstract}
Super-resolution is a machine-learning technique in image processing which generates high-resolution images from
low-resolution images.
Inspired by this approach, we perform a numerical experiment of quantum machine learning,
which takes low-resolution (low plane-wave energy cutoff) one-particle molecular wavefunctions in plane-wave basis
as input and generates high-resolution (high plane-wave energy cutoff) wavefunctions in fictitious one-dimensional
systems, and study the performance of different learning models.
We show that the trained models can generate
wavefunctions having higher fidelity values with respect to the ground-truth wavefunctions than
a simple linear interpolation, and
the results can be improved both qualitatively and quantitatively
by including data-dependent information in the ansatz.
On the other hand, the accuracy of the current approach deteriorates for wavefunctions calculated
in electronic configurations not included in the training dataset. We also discuss the generalization of this approach
to many-body electron wavefunctions.
\end{abstract}

\maketitle

\section{introduction}
Quantum machine learning (QML) is considered to be one of the potential promising applications of
near-term and future quantum computing~\cite{Schuld2015, Biamonte2017}.
Hybrid classical-quantum QML algorithms with a parameterized quantum circuit (PQC)~\cite{Cerezo2021}
are extensively studied
as a practical application of noisy intermediate-scale quantum (NISQ) devices, and
fully quantum approaches, such as quantum support vector machines~\cite{Rebentrost2014}, have also been proposed.

Although significant progress has been made in this field,
it is still unclear if QML applied on classical input data can offer quantum advantage,
except for a few special cases~\cite{Liu2021, Huang2021, Schuld2022, Kuebler2021, Qian2021}.
A new direction in QML is to use quantum states directly as input data~\cite{Perrier2021, Schatzki2021}.
This approach is most straightforwardly applicable to problems in physical science if quantum states of physical objects are encoded as input data into QML.
Many-body wavefunctions are a representative input  for classification
of topological or magnetic phases~\cite{Cong2019, Uvarov2020,  Banchi2021, Wrobel2021}, and
molecular electronic wavefunctions are also used~\cite{Romero2017, Bilkis2022}.
While encoding of the classical input data into quantum states is a highly nontrivial problem
in QML~\cite{Schuld2019, SupervisedLearning},
no such issue arises when learning directly on quantum data. In this case, the performance of a QML model
is solely determined by the structure of the ansatz, i.e., the arrangement of the parameterized unitaries, to be employed.

In this work, we consider a different type of QML experiment with quantum input
using one-particle molecular electronic wavefunctions,
inspired by the super-resolution technique developed in the field of (classical) machine learning.
Through this experiment, we study how we can improve QML models with quantum input.
As a possible improvement for QML model, we explore the possibility of
adding classical information to quantum input data in a scheme similar to
the data re-uploading~\cite{Salinas2020, Vidal2020, Schuld2021}.
The single-image super-resolution is an image processing technique
which generates a high-resolution (HR) image from a low-resolution (LR) image~\cite{Yang2019, Bashir2021}.
Existing works on super-resolution are based on several deep learning approaches, such as
deep convolutional networks~\cite{Dong2014} and generative adversarial networks~\cite{Ledig2017}.
A potential future application of this work is to generate approximate electron wavefunctions in
high-throughput quantum chemistry calculations with low computational cost.

We consider molecular one-particle
wavefunctions expressed in plane waves. The plane-wave basis is a flexible basis widely used
in the electronic structure calculations of solid-state systems~\cite{Martin2004}, and several
quantum algorithms based on plane-wave expansion of the wavefunctions
have recently been proposed for fault-tolerant quantum computers, both in the
second and first quantization formalisms~\cite{Berry2018, Babbush2018, Babbush2019, Su2021, OBrien2021, Delgado2022}.
In this work, ``low-resolution'' wavefunctions are expressed with a small number of plane waves (low plane-wave energy cutoff),
and by using them as input of QML
we aim to construct ``high-resolution'' wavefunctions which are expressed
with a larger number of plane waves.
As a benchmark test of this approach,
we consider fictitious molecular systems in one spatial dimension~\cite{Baker2015}.

The state of an electronic system with a fixed number of particles can be encoded into a quantum register
through either the first or second quantization approaches.
The first option may use significantly fewer qubits, especially when the number of particles is small,
while the second option uses as many qubits as the number of basis functions.
Because our one-particle wavefunctions are expanded into up to 31 plane-waves,
we opt for the first quantization encoding.
The price we have to pay for this dense encoding in first quantization
is that the resulting quantum states are highly entangled. In our experiments, we use
exact amplitude encoding for the LR/HR wavefunctions
and use rather deep parameterized circuits with a statevector-based quantum simulator.
Simplification of the ansatzes and consideration of various runtime errors, which would be required
for this work to be practically applicable in the NISQ era, are left for future research.
We also discuss a generalization of this approach to many-body wavefunctions.

\section{method}
\subsection{Molecular orbitals in plane-wave basis}
We consider fictitious one-dimensional molecular systems interacting
via exponential Coulomb-mimicking functions proposed in Ref.~\cite{Baker2015}, although the current approach
can be generalized to treat general three-dimensional systems.
We impose a periodic boundary condition with periodicity $L$
and expand the one-particle wavefunctions (molecular orbitals) $\psi_{\mu\sigma}(x)$
with $N_{\textrm{pw}}$ plane-wave basis functions $\chi_{k_{j}}(x)$ as
\begin{equation}
\psi_{\mu\sigma}(x)=\sum_{k_{j}} C_{k_{j}\mu\sigma}\chi_{k_{j}}(x),
\end{equation}
where $\mu$ and $\sigma$ are spatial orbital and spin indices, respectively, and
\begin{equation}
    \chi_{k_{j}}(x) = \frac{1}{\sqrt{L}} e^{i k_{j} x},
\end{equation}
\begin{equation}
    k_{j} = \frac{2 \pi}{L} j \quad j = 0, \pm 1, \pm 2, \dots, \pm\frac{N_{\textrm{pw}} - 1}{2}.
    \label{eq:k}
\end{equation}
We consider odd $N_{\textrm{pw}}$. The expansion coefficients $C_{k_{j}\mu\sigma}$ are determined
by solving the following Hartree-Fock equation self-consistently
\begin{equation}
    \sum_{k_{j'}}h^{\textrm{HF}\sigma}_{jj'} C_{k_{j'}\mu\sigma} = \varepsilon_{\mu\sigma} C_{k_{j}\mu\sigma},
    \label{eq:hf}
\end{equation}
where $h^{\textrm{HF}}$ is the Hartree-Fock one-particle Hamiltonian matrix and $\varepsilon_{\mu\sigma}$ are Hartree-Fock
eigenenergies.
The details of the calculations are given in Appendix~\ref{app:pw}.

\subsection{Quantum circuit for resolution enhancement}
\begin{figure*}[tbp]
\centering
\includegraphics[clip]{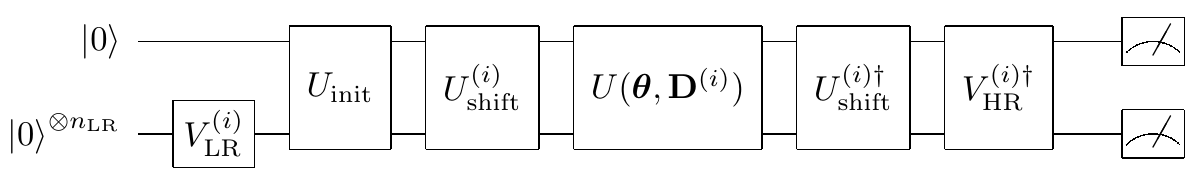}
\caption{Quantum circuit to calculate the fidelity between the $i$-th
true HR wavefunction and the predicted wavefunction generated from the corresponding LR data.}
\label{fig:qc_fidelity}
\end{figure*}
\begin{figure}[tbp]
\centering
\includegraphics[clip]{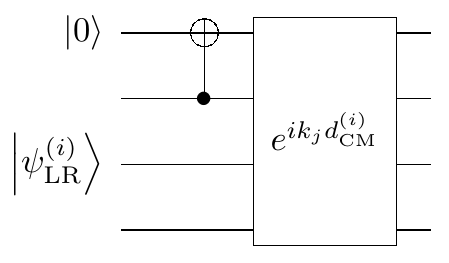}
\caption{The initialization operation $U_{\textrm{init}}$ (first CNOT gate)
and the shift operation $U^{(i)}_{\textrm{shift}}$ ($e^{i k_{j} d^{(i)}_{\textrm{CM}}}$) in
Fig.~\ref{fig:qc_fidelity} for $n_{\textrm{LR}}=3$.
Here $k_{j}$ are plane-wave wavevectors,
and $d^{(i)}_{\textrm{CM}}$ is the center of mass for the $i$-th data.}
\label{fig:qc_uinit}
\end{figure}

We encode the LR molecular wavefunction coefficients $C^{\textrm{LR}}_{k_{j}\mu\sigma}$
for $N^{\textrm{LR}}_{\textrm{pw}}$ plane waves
in a quantum circuit as the amplitudes of $n_{\textrm{LR}} = \lceil\log_{2}N_{\textrm{pw}}^{\textrm{LR}}\rceil$ qubit states,
and
generate the HR wavefunction coefficients  $C^{\textrm{HR}}_{k_{j}\mu\sigma}$
for $N^{\textrm{HR}}_{\textrm{pw}} (> N^{\textrm{LR}}_{\textrm{pw}})$ plane waves
with $n_{\textrm{HR}} = \lceil\log_{2}N_{\textrm{pw}}^{\textrm{HR}}\rceil$ qubits,
using parametrized quantum circuits whose parameters are determined by minimizing some cost function.
We consider the case $n_{\textrm{HR}}=n_{\textrm{LR}} + 1$, although more general resolution enhancement
($n_{\textrm{HR}} = n_{\textrm{LR}} + m, m \in \mathbb{N}$) is also possible.
We consider the case where the LR wavefunctions are underconverged. Under this condition,
the HR wavefunctions cannot be obtained by a simple interpolation of the corresponding LR wavefunctions
in real space, but have to be obtained by the extrapolation of the LR wavefunction coefficients $C_{k_{j}\mu\sigma}$
in the Fourier (wavevector) space.
We prepare LR/HR wavefunction pairs $\{\psi_{\text{LR/HR}}^{(i)}\}_{i=1}^{N_{\text{data}}}$ as datasets
and try to find $U(\bm{\theta},\bm{D}^{(i)})$
which can generate approximate HR wavefunctions~$|\tilde{\psi}_{\text{HR}}^{(i)}\rangle$
 from LR wavefunctions as
$|\tilde{\psi}_{\text{HR}}^{(i)}\rangle=U(\bm{\theta},\bm{D}^{(i)})|0\rangle\otimes|\psi^{(i)}_{\text{LR}}\rangle$
 by optimizing parameters $\theta$.
Here $i$ is the sample index.
The ansatz $U(\boldsymbol\theta, \mathbf{D}^{(i)})$
may also depend on the data-dependent parameters $\mathbf{D}^{(i)}$ for the sample $i$.
The state $|\psi^{(i)}_{\textrm{LR}}\rangle$ is the $i$-th LR wavefunction expressed using $n_{\textrm{LR}}$ qubits.
Unitary operations guarantee that LR plane-wave basis wavefunctions are mapped to orthonormal HR wavefunctions.
The parameters $\boldsymbol\theta$
are optimized by maximizing the fidelity between the predicted and the true HR orbitals,
or equivalently by minimizing the cost function
\begin{equation}
    \mathcal{L} = -\frac{1}{N_{\textrm{data}}}\sum_{i=1}^{N_{\textrm{data}}} |\langle \psi^{(i)}_{\textrm{HR}}|\tilde{\psi}^{(i)}_{\textrm{HR}}\rangle|^{2}.
    \label{eq:cost}
\end{equation}

Figure~\ref{fig:qc_fidelity} shows the quantum circuit used
to compute the fidelity between $|\psi^{(i)}_{\textrm{HR}}\rangle$
and $|\tilde{\psi}^{(i)}_{\textrm{HR}}\rangle$. Here $V_{\textrm{LR}}^{(i)}$ and $V_{\textrm{HR}}^{(i)}$ are
unitary operations to encode $C^{\textrm{LR}(i)}_{k_{j}\mu\sigma}$ and $C^{\textrm{HR}(i)}_{k_{j}\mu\sigma}$,
respectively.
We encode the LR wavefunctions as 
\begin{equation}
|\psi^{(i)}_{\textrm{LR}}\rangle = V_{\textrm{LR}}^{(i)}|0\rangle^{\otimes n_{\textrm{LR}}}
=\sum_{j=-\frac{N^{\textrm{LR}}_{\textrm{pw}} - 1}{2}}^{+\frac{N^{\textrm{LR}}_{\textrm{pw}} - 1}{2}}C^{\textrm{LR}(i)}_{k_{j}\mu\sigma}|k_{j}\rangle,
\end{equation}
where $|k_{j}\rangle$ are expressed by following
the convention in the Fast Fourier Transform~\cite{NumericalRecipes}
\begin{equation}
    |k_{j}\rangle = \left\{
    \begin{array}{ll}
        |j\rangle & (j \ge 0) \\
        |j + 2^{n_{\textrm{LR}}}\rangle & (j < 0)
    \end{array}
    \right.
\end{equation}
with $|j\rangle$ the binary representation of $j$.
The HR wavefunctions are similarly embedded in the HR space with $n_{\textrm{HR}}=n_{\text{LR}}+1$ qubits with
$V_{\textrm{HR}}^{(i)}$.
The two unitaries $U_{\textrm{init}}$ and $U^{(i)}_{\textrm{shift}}$
in Fig.~\ref{fig:qc_fidelity}
are the initialization and shifting operations, respectively, which are described below.
The fidelity is obtained as the probability of observing $|0\rangle^{\otimes n_{\textrm{HR}}}$ in the measurements in
Fig.~\ref{fig:qc_fidelity}.

Figure~\ref{fig:qc_uinit} shows the initialization part of Fig.~\ref{fig:qc_fidelity}.
The first CNOT reorders the coefficients $C^{\textrm{LR}(i)}_{k_{j}\mu\sigma}$
in the $2^{n_{\textrm{LR}} + 1}$-dimensional space, and
the second operation multiplies a phase factor $e^{i k_{j} d^{(i)}_{\textrm{CM}}}$ to each of the plane-wave basis, which
corresponds to setting the origin to $d_{\textrm{CM}}^{(i)}$, the center of mass of sample $i$.

\subsection{Dataset and ansatz}
\begin{figure}[tbp]
\centering
\includegraphics[clip]{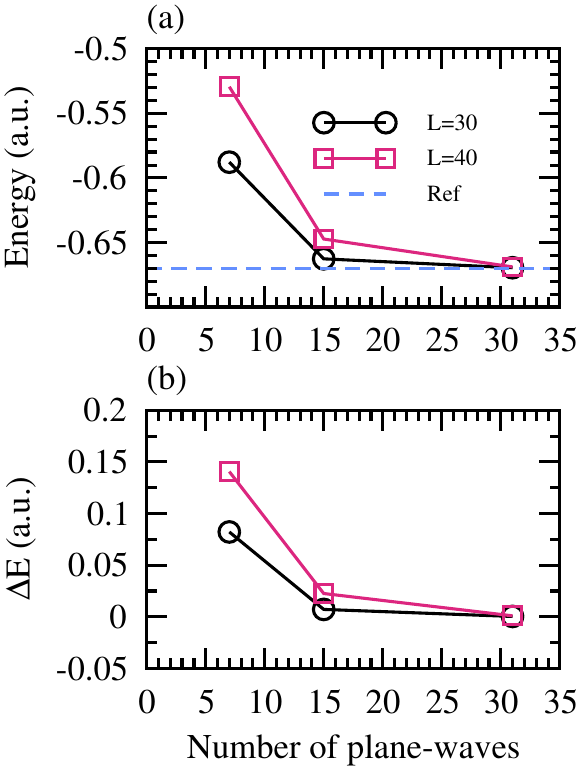}
\caption{(a) The calculated energy of a hydrogen atom for $L=30$ a.u. (circles) and $L=40$ a.u. (squares) with
$N_{\textrm{pw}} = 7, 15, 31$. The result of a real space grid approach
in Ref.~\cite{Baker2015} ($E_{\textrm{ref}}=-0.670$) is also shown as a dashed line for comparison.
(b) The energy difference defined as $|E - E_{\textrm{ref}}|$.}
\label{fig:h_np}
\end{figure}
As a benchmark test of the approach, we consider the wavefunctions of
the systems consisting of hydrogen atoms, H$_{x}$, similar to those used in Ref.~\cite{Li2021}.
We consider only symmetric molecules and place them so that their centers are at $x=0$. With this choice,
the wavefunction coefficients $C_{k_{j}\mu\sigma}$ can be chosen to be real, and
the shifting operation $U^{(i)}_{\textrm{shift}}$ in Fig.~\ref{fig:qc_fidelity} is not required.
We use 52 Hartree-Fock occupied wavefunctions for training, and the trained model is applied
to the validation dataset, which includes molecules not included in the training dataset
and uses a finer bond length grid. The details of the training and validation datasets are described
in Appendix \ref{app:dataset}.

Two cases are considered in our experiments: (i) LR (HR) wavefunctions are constructed with
$N_{\textrm{pw}}=7\ (15)$ plane waves at $L=30$ a.u.;
(ii) LR (HR) wavefunctions are constructed with $N_{\textrm{pw}}=15\ (31)$ plane waves at $L=40$ a.u..
The number of qubits used in our experiments is four and five for cases (i) and (ii), respectively.
The total energies of a hydrogen atom calculated with these parameters are shown in Figs.~\ref{fig:h_np} (a, b).
In case (i)
($N_{\textrm{pw}}: 7\to 15$)
the Fourier wavevector grid of the LR wavefunctions is very coarse,
making the LR wavefunctions very crude approximations,
while in case (ii) ($N_{\textrm{pw}}: 15\to 31$)
the grid is finer and the LR wavefunctions are closer to the exact wavefunctions.
With $N_{\textrm{pw}}=31$, the calculated energy values are within 0.01 a.u.~of the reference energy
obtained from a real space grid approach in Ref.~\cite{Baker2015}
for both $L=30$ and $L=40$.

\begin{figure}
    \centering
    \includegraphics[width=70mm]{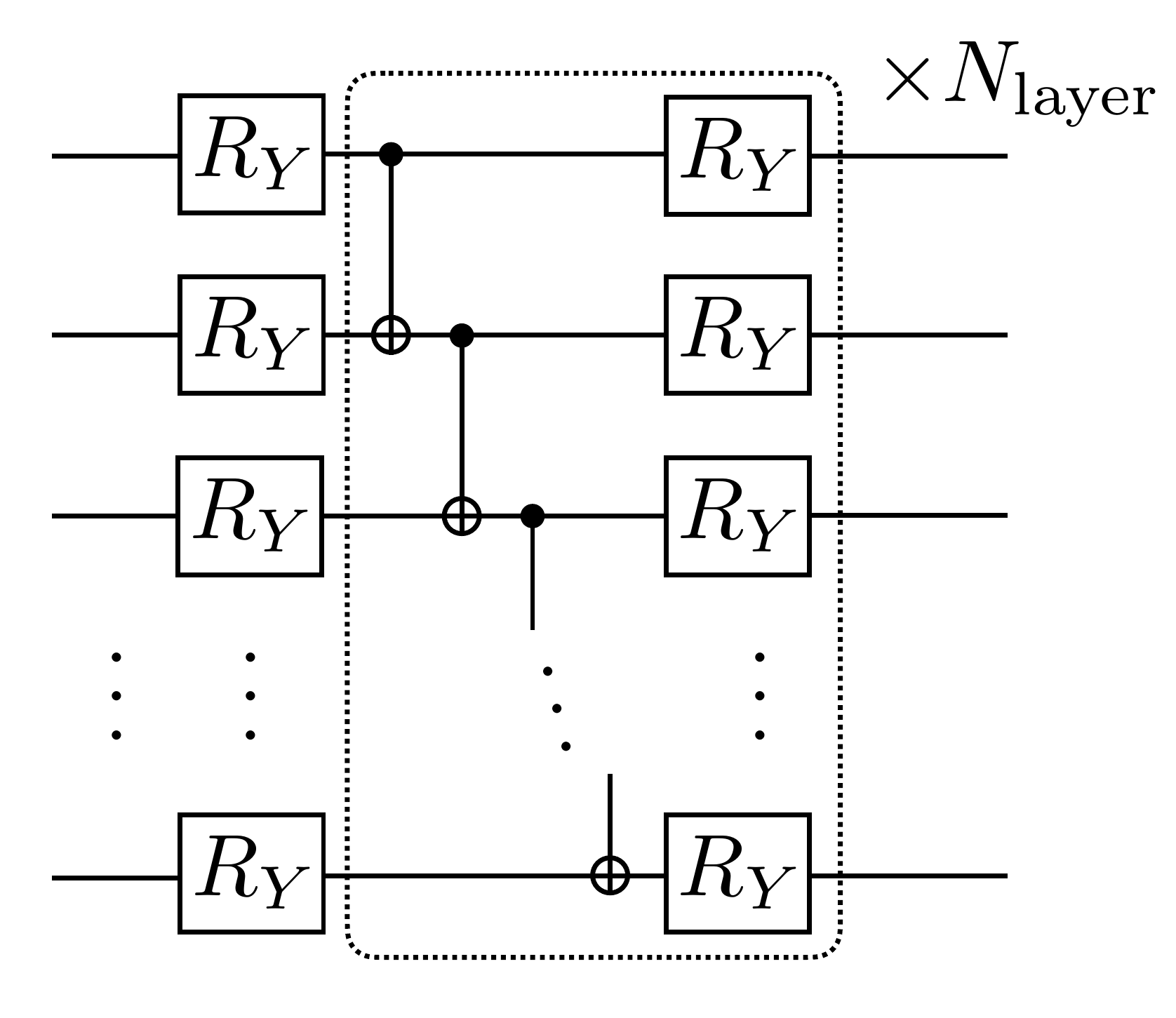}
    \caption{Circuit diagram of ansatz 1.}
    \label{fig:ansatz1}
\end{figure}

Two classes of ansatzes are employed as the unitary $U(\boldsymbol\theta, \mathbf{D}^{(i)})$:
the first (ansatz 1) is a general ansatz for real amplitudes
and is independent of the input sample,
\begin{equation}
    U(\boldsymbol\theta) =
    \prod_{l=1}^{N_{\textrm{layer}}}
    \Bigl [
    \Bigl (
    \prod_{i_{q}=1}^{n_{q}} R^{i_{q}}_{y}(\theta_{i_{q}, l})
    \Bigr )
     U_{\textrm{Ent}}
    \Bigr ]
    \prod_{i_{q}=1}^{n_{q}} R^{i_{q}}_{y}(\theta_{i_{q}, 0}),
    \label{eq:ansatz1}
\end{equation}
where $N_{\textrm{layer}}$ and $n_{q}$ are the number of layers and qubits, respectively,
and $U_{\textrm{Ent}}$ is an entangling operator consisting of $n_{q} - 1$ CNOT gates connecting the $i$-th and
$(i+1)$-th qubits. The total number of parameters in this class of ansatzes is $n_{q}(N_{\textrm{layer}} + 1)$.
See~Fig.~\ref{fig:ansatz1} for a circuit diagram.

\begin{figure}
\begin{minipage}{0.8\linewidth}
\centering
\includegraphics[width=70mm]{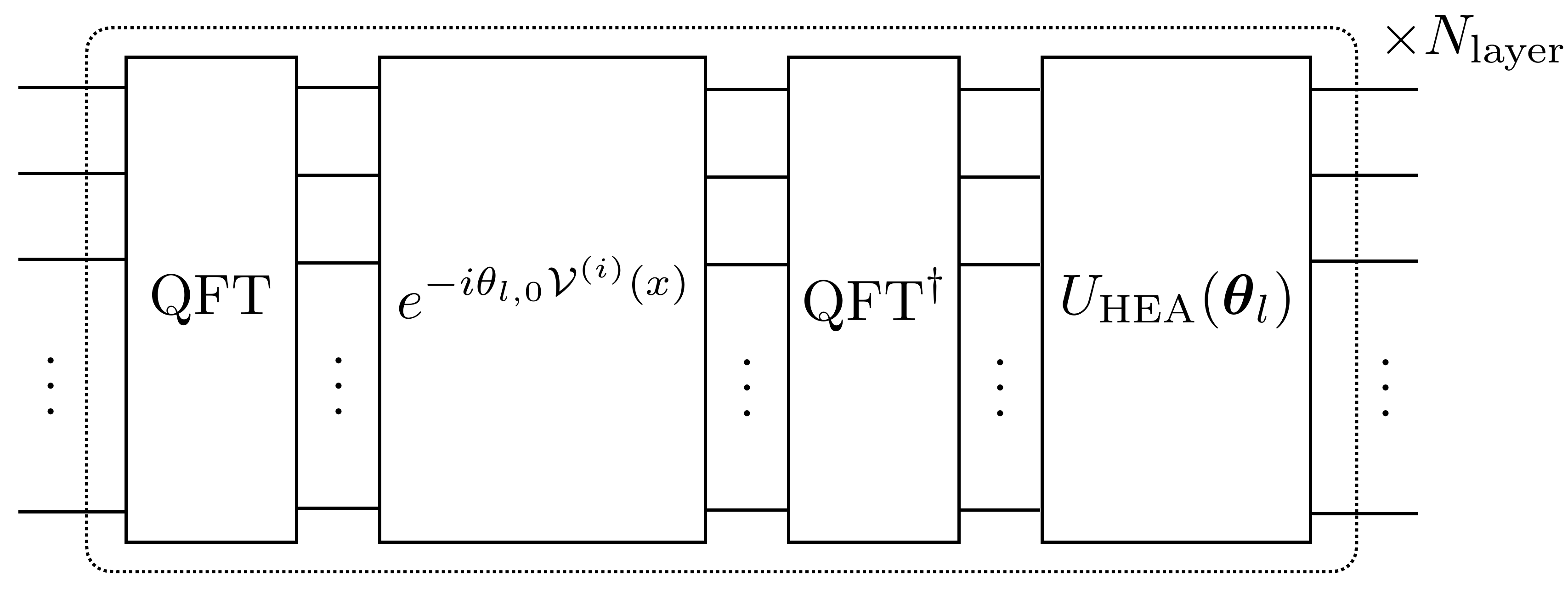}
\subcaption{full ansatz}
\end{minipage}
\\
\begin{minipage}{0.8\linewidth}
\centering
\includegraphics[width=70mm]{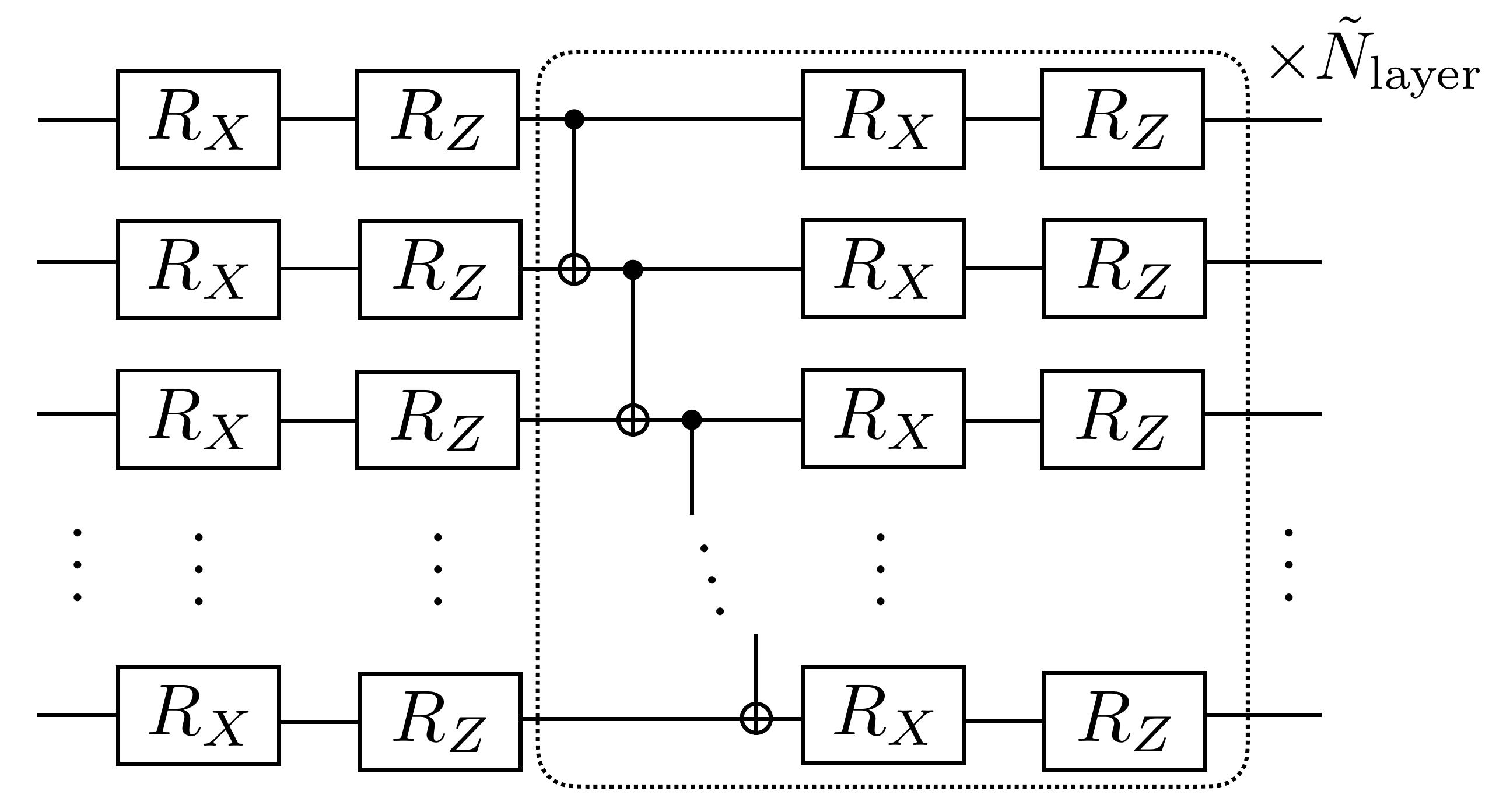}
\subcaption{HEA part}
\end{minipage}
\caption{Circuit diagram of ansatz 2.}
\label{fig:ansatz2}
\end{figure}

The second class (ansatz 2) consists of sample-dependent ansatzes
inspired by the Fourier-transformed variational Hamiltonian ansatz~\cite{Choquette2021}
\begin{equation}
    U(\boldsymbol\theta, \mathbf{D}^{(i)}) = \prod_{l=1}^{N_{\textrm{layer}}} U_{\textrm{HEA}}(\boldsymbol\theta_{l})
    \textrm{QFT}^{\dagger} e^{-i \theta_{l, 0} \mathcal{V}^{(i)}(x)} \textrm{QFT},
    \label{eq:ansatz2}
\end{equation}
where $U_{\textrm{HEA}}$ is a hardware-efficient ansatz
\begin{eqnarray}
U_{\textrm{HEA}}(\boldsymbol\theta_{l}) &=&
\prod_{\tilde{l}=1}^{\tilde{N}_{\textrm{layer}}}
\Bigl [
    \Bigl (
    \prod_{i_{q}=1}^{n_{q}} R^{i_{q}}_{z}(\theta_{l, i_{q}, \tilde{l}, z})R^{i_{q}}_{x}(\theta_{l, i_{q}, \tilde{l}, x})
    \Bigr )
    U_{\textrm{Ent}}
    \Bigr ] \nonumber\\
    && \times
    \prod_{i_{q}=1}^{n_{q}} R^{i_{q}}_{z}(\theta_{l, i_{q}, 0, z}) R^{i_{q}}_{x}(\theta_{l, i_{q}, 0, x}),
    \label{eq:uhea}
\end{eqnarray}
with $\tilde{N}_{\textrm{layer}}$ the number of sublayers.
The corresponding circuit diagram is shown in Fig.~\ref{fig:ansatz2}.
The first QFT in Eq.~(\ref{eq:ansatz2}) performs the Quantum Fourier Transform
and transforms the input LR wavefunction from the Fourier space to real space,
$|k_{j}\rangle \to |x_{j}\rangle$. The following operator adds phase factors
to the states as $|x_{j}\rangle \to e^{-i\theta_{l,0} \mathcal{V}^{(i)}(x_{j})}|x_{j}\rangle$, where $\mathcal{V}^{(i)}(x)$ is
a sample-dependent effective one-particle potential. While $\mathcal{V}^{(i)}(x)$ can be any function of the electron position $x$ in principle,
we use the electron-nuclei Coulomb potential $v_{\textrm{en}}(x)$ in Eq.~(\ref{eq:el-nuc}),
which depends only on the numbers and positions of the atoms in each sample, as the simplest choice.
The wavefunction is transformed back to the Fourier space with the next QFT$^{\dagger}$
before the last operation $U_{\textrm{HEA}}(\boldsymbol\theta_{l})$,
which accounts for the kinetic part of the Hamiltonian and also
other effects not included in $e^{-i\theta_{l,0}\mathcal{V}^{(i)}(x)}$.

We choose the number of layers $N_{\textrm{layer}}=32$ for ansatz~1 (Eq.~(\ref{eq:ansatz1})). We use this
large number of layers to provide this model with sufficient flexibility.
We observe that increasing the number of layers up to $N_{\textrm{layer}}=64$ does not improve the results significantly.
For ansatz 2 (Eq.~(\ref{eq:ansatz2})), $N_{\textrm{layer}}=3, \tilde{N}_{\textrm{layer}}=8$ are used.
The number of variational parameters in ansatz 1 (ansatz 2) are  132 (219) and 165 (273) for case (i) and case (ii),
respectively.
Because of the data-dependent potential term, ansatz 2 is more easily trapped in local minima than ansatz 1.
We find that starting with $\boldsymbol\theta = 0$ in ansatz 2 generally shows a good convergence.
We use the PennyLane library~\cite{PennyLane} and minimize the cost function (Eq.~(\ref{eq:cost}))
using the Adam optimizer.

\subsection{Extension to many-body wavefunctions}
\begin{figure*}[tbp]
\centering
\includegraphics[clip]{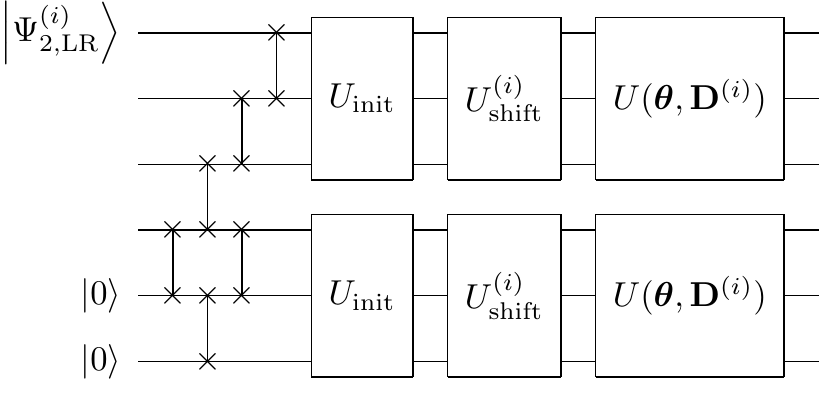}
\caption{Quantum circuit to generate a HR two-electron wavefunction from the $i$-th LR wavefunction
    $\Psi^{(i)}_{2,\textrm{LR}}$
    (represented by the first four qubits of the circuit) for $n_{\textrm{LR}}=2$.}
\label{fig:qc_2el}
\end{figure*}
In this section we consider whether we can apply the model trained with one-particle wavefunctions directly
to enhance the resolution of many-body wavefunctions.
Neglecting spin degrees of freedom for simplicity, in first quantization,
the $N_{\textrm{el}}$-electron wavefunctions can be written with
$N_{\textrm{el}} \lceil \log_{2} N_{\textrm{pw}} \rceil$ qubits
as~\cite{Su2021}
\begin{equation}
    |\Psi_{N_{\textrm{el}}}\rangle =
    \sum_{\{k_{j}\}} C_{k_{0}k_{1}\dots k_{N_{\textrm{el}}-1}} |k_{0} k_{1} \dots k_{N_{\textrm{el}}-1}\rangle,
    \label{eq:phin}
\end{equation}
where the expansion coefficients are antisymmetric:
\begin{equation}
C_{\dots k_{i}\dots k_{j}\dots} = -C_{\dots k_{j}\dots k_{i}\dots}.
\end{equation}
When $U(\boldsymbol \theta, \mathbf{D}^{(i)})$ does not contain orbital-dependent
parameters, as in ansatz 1 and ansatz 2 in the previous section,
one can generate an approximate antisymmetric HR many-body wavefunction from a LR many-body state expressed with
$N_{\textrm{el}} n_{\textrm{LR}}$ qubits and with extra $N_{\textrm{el}}$ qubits,
$|\Psi^{\textrm{LR}(i)}_{N_{\textrm{el}}}\rangle \otimes |0\rangle^{\otimes N_{\textrm{el}}}$,
by applying $\bar{U}^{(i)} = U(\boldsymbol\theta,  \mathbf{D}^{(i)})U_{\textrm{shift}}^{(i)} U_{\textrm{init}}$
to each of the $N_{\textrm{el}}$ registers as
\begin{eqnarray}
    |\tilde{\Psi}^{\textrm{HR}(i)}_{N_{\textrm{el}}}\rangle &\approx&
    \Bigl [
    \bar{U}^{(i)} \otimes \bar{U}^{(i)} \otimes \cdots \otimes
    \bar{U}^{(i)}
    \Bigr ] \nonumber\\
    &&
    U_{\textrm{SWAP}}
    |\Psi^{\textrm{LR}(i)}_{N_{\textrm{el}}}\rangle \otimes |0\rangle^{\otimes N_{\textrm{el}}}.
    \label{eq:phinel}
\end{eqnarray}
Here, $U_{\textrm{SWAP}}$ is a unitary operation that reorders
states $|k_{0}k_{1}\dots k_{N_{\textrm{el}}-1}\rangle\otimes |0\rangle^{\otimes N_{\textrm{el}}}$ to
$|0 \, k_{0} 0\, k_{1}\dots 0\, k_{N_{\textrm{el}}-1}\rangle$ using SWAP gates.
The case for $N_{\textrm{el}}=2$ is shown in Fig.~\ref{fig:qc_2el}.
Spin degrees of freedom can be included in Eq.~(\ref{eq:phinel}) by adding extra $N_{\textrm{el}}$ qubits.

Although this procedure is simple and very easy to implement,
there are two potential problems in this generalization.
First,
since a many-body wavefunction is written as a sum of the
product of $N_{\textrm{el}}$ one-particle wavefuntion in first quantization, the fidelity
between the predicted and true HR wavefunctions,
$|\langle \tilde{\Psi}^{\textrm{HR}}_{N_{\textrm{el}}}|\Psi^{\textrm{HR}}_{N_{\textrm{el}}}\rangle|^{2}$, decreases as
$O\bigl( (f_{1})^{N_{\textrm{el}}} \bigr)$, where $f_{1}$ is a typical fidelity
of one-particle wavefunctions obtained with the ansatz used.
A scalable generalization of this approach to many-body systems would require some correction operation that
entangles qubits in different registers
and that still keeps the antisymmetry of the wavefunctions. This point is left for our future research.
The second problem is the phase of the predicted wavefunctions;
in the current approach based on fidelity maximization, the unitary operation $\bar{U}^{(i)}$ can add
orbital-dependent phase factors to the predicted one-particle wavefunctions as
\begin{equation}
\bar{U}^{(i)}\psi^{\textrm{LR}}_{\mu\sigma} \approx  e^{i\theta_{\mu}^{(i)}}\psi^{\textrm{HR}}_{\mu\sigma}.
\label{eq:phase}
\end{equation}
This implies that each term in the predicted wavefunction in Eq.~(\ref{eq:phin}) can potentially
get a different relative phase, resulting in a low-fidelity state.
A simple fix to this problem is to consider a linear combination of orbital pairs,
\begin{equation}
    \psi^{(i)}_{\mu_{1}\mu_{2}\sigma} =
    \frac{1}{\sqrt{2}}
    \Bigl [
    \psi^{(i)}_{\mu_{1}\sigma} + \psi^{(i)}_{\mu_{2}\sigma}
    \Bigr ],
    \label{eq:2orb}
\end{equation}
and include $\psi^{(i)}_{\mu_{1}\mu_{2}\sigma}$ in the training dataset,
which has an effect of aligning the two phases $e^{i\theta_{\mu_{1}}}$ and
$e^{i\theta_{\mu_{2}}}$ in Eq.~(\ref{eq:phase}).
In addition, the phases of LR and HR wavefunctions in the training
dataset must also be consistent. This can be done by preprocessing the training dataset, as described
in Appendix \ref{app:dataset}.

\section{results and discussion}
\subsection{One-electron case}
\begin{table}
    \begin{ruledtabular}
        \begin{tabular}{r c c}
            & case (i) & case (ii) \\
            & $N_{\textrm{pw}}:7\to 15, L=30$ & $N_{\textrm{pw}}:15\to31, L=40$ \\ \hline
            no ansatz            & 0.890 & 0.981 \\
            ansatz 1             & 0.959 & 0.992 \\
            ansatz 2             & 0.983 & 0.997 \\
            linear interpolation & 0.839 & 0.954
        \end{tabular}
        \caption{Calculated average fidelities of the training dataset.}
        \label{tbl:fidelity}
    \end{ruledtabular}
\end{table}
Table ~\ref{tbl:fidelity} shows the calculated average fidelity values between the ground-truth and predicted HR wavefunctions
for the training dataset.
In the table we also show the results of the linear interpolation in real space calculated with one ancilla qubit,
which is detailed in Appendix \ref{app:intp}.
In case (i) ($N_{\textrm{pw}}: 7\to 15$), as the LR wavefunctions are very crude approximations of the HR
wavefunctions, without including any ansatz we get a low average fidelity of 0.890.
The linear interpolation in real space, shown in the last row in Table~\ref{tbl:fidelity},
performs even poorer.
Of the two parameterized models, ansatz 2 gives better results.
This may owe to the nonlinearity introduced in the model through the inclusion of sample-dependent information
in the ansatz. We further note that this information is inserted into the ansatz repeatedly,
which may improve the expressibility of the ansatz in a similar manner to the data reuploading technique
used to encode classical information into quantum states~\cite{Salinas2020}.
In case (ii) ($N_{\textrm{pw}}: 15 \to 31$) the LR wavefunctions already have
high fidelity values with the HR wavefunctions,
but the results are improved also in this case by including the two ansatzes.

\begin{figure}[tbp]
\centering
\includegraphics[clip]{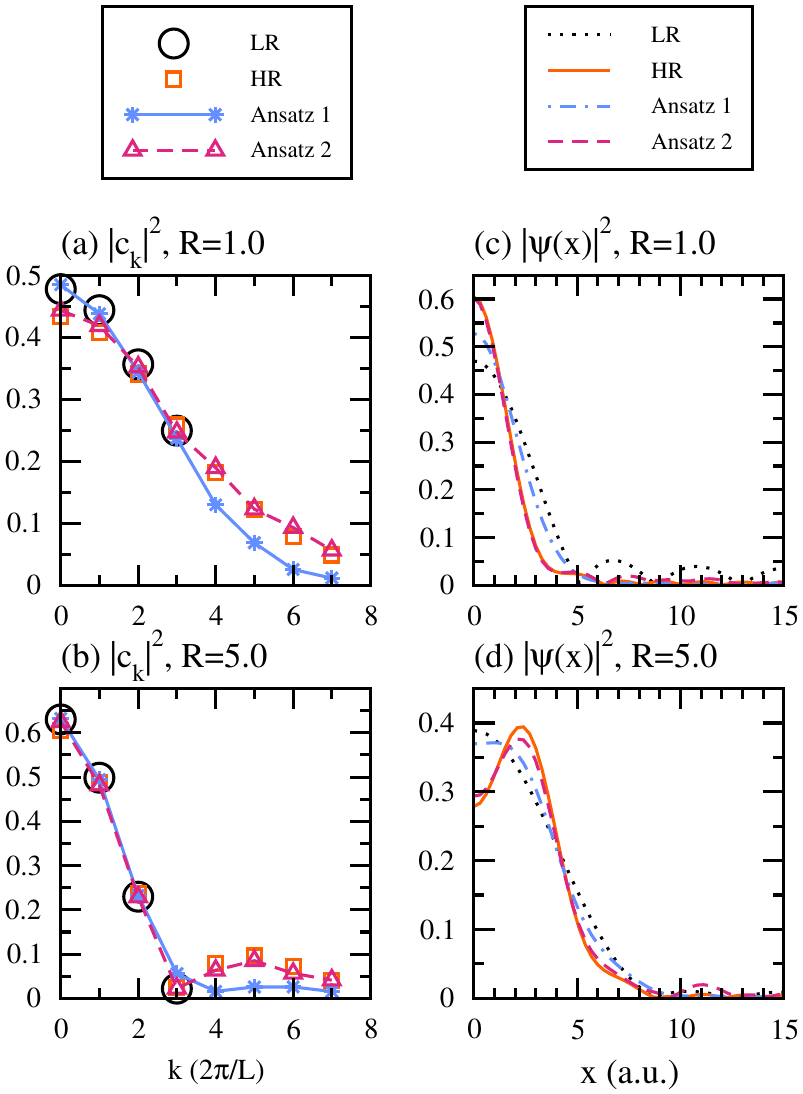}
\caption{The occupied Hartree-Fock wavefunction of the singlet ground state of
H$_{2}$ at bond length $R=1.0$ a.u.~(a, c) and $R=5.0$ a.u.~(b, d) in Fourier space (a, b) and in real space (c, d)
at $L=30$ a.u.~for $N_{\textrm{pw}}:7 \to 15$ resolution enhancement.
The hydrogen atoms are located at $x=\pm \frac{R}{2}$. Only $k\ge 0$ and $0 \le x\le \frac{L}{2}$ regions are shown.}
\label{fig:wfn_h2_l30_07-15_singlet}
\end{figure}
\begin{figure}[tbp]
\centering
\includegraphics[clip]{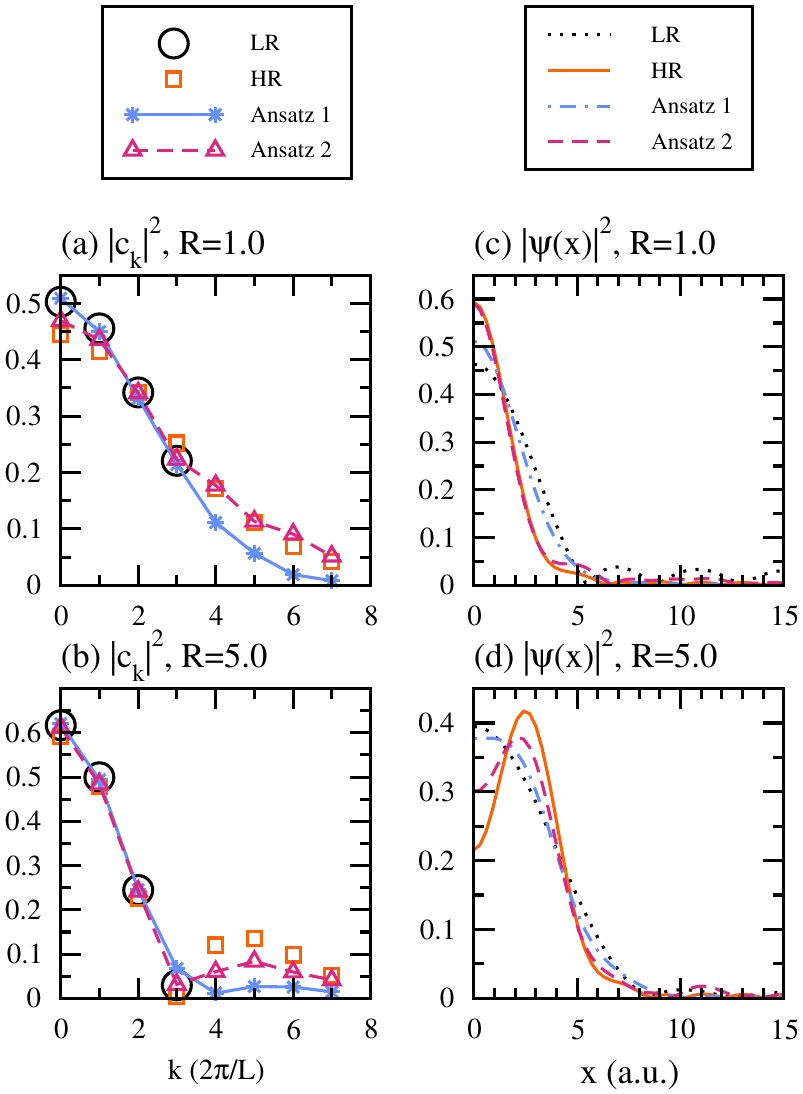}
\caption{The lowest occupied Hartree-Fock wavefunction of the triplet ground state of
H$_{2}$ at bond length $R=1.0$ a.u.~(a, c) and $R=5.0$ a.u.~(b, d)
at $L=30$ a.u.~for $N_{\textrm{pw}}:7 \to 15$ resolution enhancement.}
\label{fig:wfn_h2_l30_07-15_triplet_1}
\end{figure}
\begin{figure}[tbp]
\centering
\includegraphics[clip]{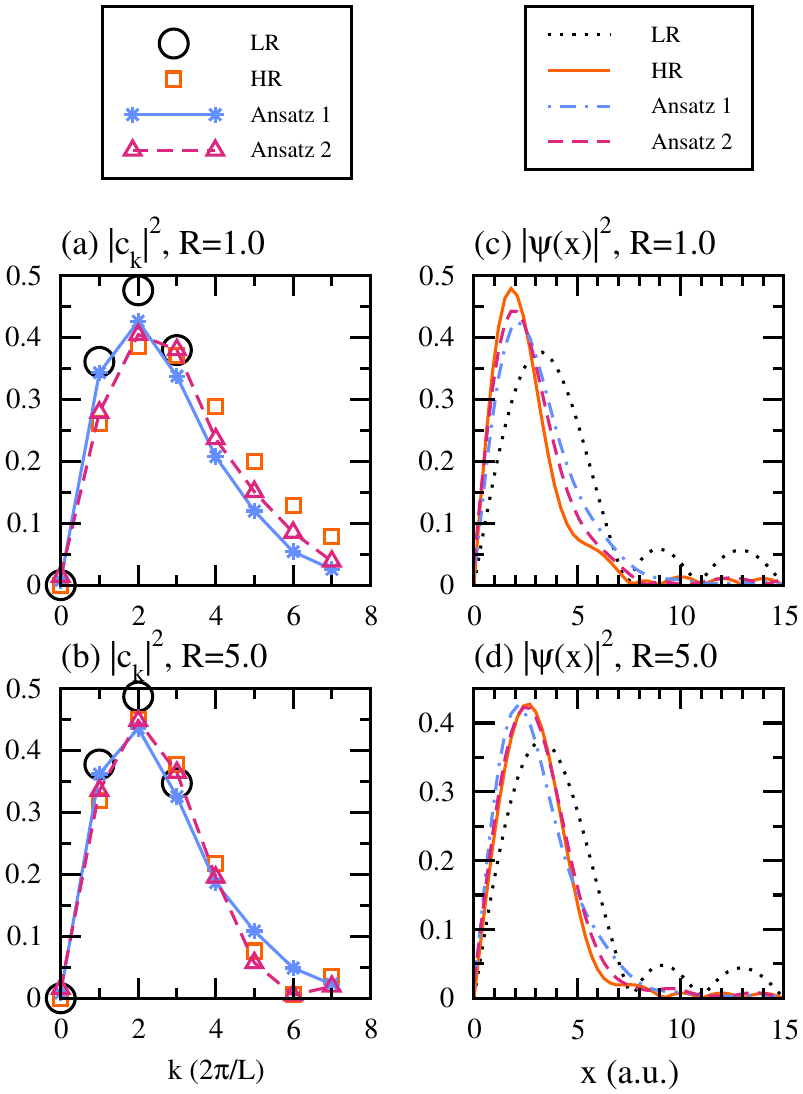}
\caption{The second lowest occupied Hartree-Fock wavefunction of the triplet ground state of
H$_{2}$ at bond length $R=1.0$ a.u.~(a, c) and $R=5.0$ a.u.~(b, d)
at $L=30$ a.u.~for $N_{\textrm{pw}}:7 \to 15$ resolution enhancement.}
\label{fig:wfn_h2_l30_07-15_triplet_2}
\end{figure}
In order to investigate the qualitative difference between the results of the two ansatzes,
in Fig.~\ref{fig:wfn_h2_l30_07-15_singlet} (a--d) we show the results of case (i) for the singlet
ground state of the H$_{2}$ molecule at two bond lengths, $R=1.0$ and $R=5.0$.
In the Fourier space (Fig.~\ref{fig:wfn_h2_l30_07-15_singlet} (a, b)), both ansatzes
extrapolate the LR wavefunction coefficients $C^{\textrm{LR}}_{k_{j}\mu}$
from the LR region ($k \le 3$), but only ansatz 2
reproduces a subpeak centered at around $k = 5$ in the $R=5.0$ result
(Fig.~\ref{fig:wfn_h2_l30_07-15_singlet} (b)).
Ansatz 1 fails to reproduce this structure, and
because of the absence of this peak in the Fourier space,
the peak of the real-space wavefunction at $x=\pm \frac{R}{2} = 2.5$
in Fig.~\ref{fig:wfn_h2_l30_07-15_singlet} (d) is not reproduced.
Similar results are obtained for the two occupied wavefunctions of the triplet H$_{2}$, shown in
Fig.~\ref{fig:wfn_h2_l30_07-15_triplet_1} and Fig.~\ref{fig:wfn_h2_l30_07-15_triplet_2}, where
ansatz 2 better approximates the ``tail'' region (i.e. $k > 3$) and
only ansatz 2 can reproduce a subpeak in Fig.~\ref{fig:wfn_h2_l30_07-15_triplet_1} (b), although the agreement is not
as good as the singlet case. The calculated
fidelity values for ansatz 1 (ansatz 2) for the singlet (Fig.~\ref{fig:wfn_h2_l30_07-15_singlet} (b))
and lowest triplet (Fig.~\ref{fig:wfn_h2_l30_07-15_triplet_1} (b)) orbitals at $R=5.0$ are
0.974 (0.996) and 0.930 (0.978), respectively.

\begin{figure}[tbp]
\centering
\includegraphics[clip]{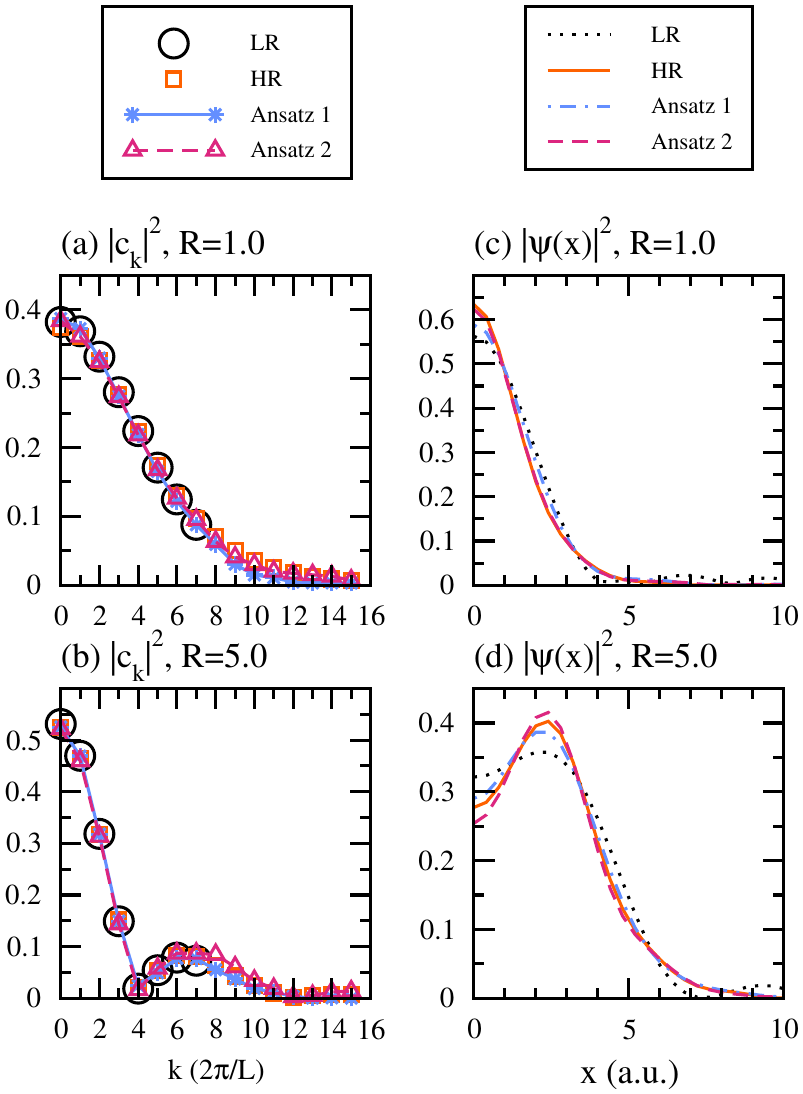}
\caption{The occupied Hartree-Fock wavefunctions of the singlet ground state of
H$_{2}$ at bond length $R=1.0$ a.u.~(a, c) and $R=5.0$ a.u.~(b, d) in Fourier space (a, b) and in real space (c, d)
with $L=40$ a.u.~for $N_{\textrm{pw}}:15 \to 31$ resolution enhancement.}
\label{fig:wfn_h2_l40_15-31_singlet}
\end{figure}
\begin{figure}[tbp]
\centering
\includegraphics[clip]{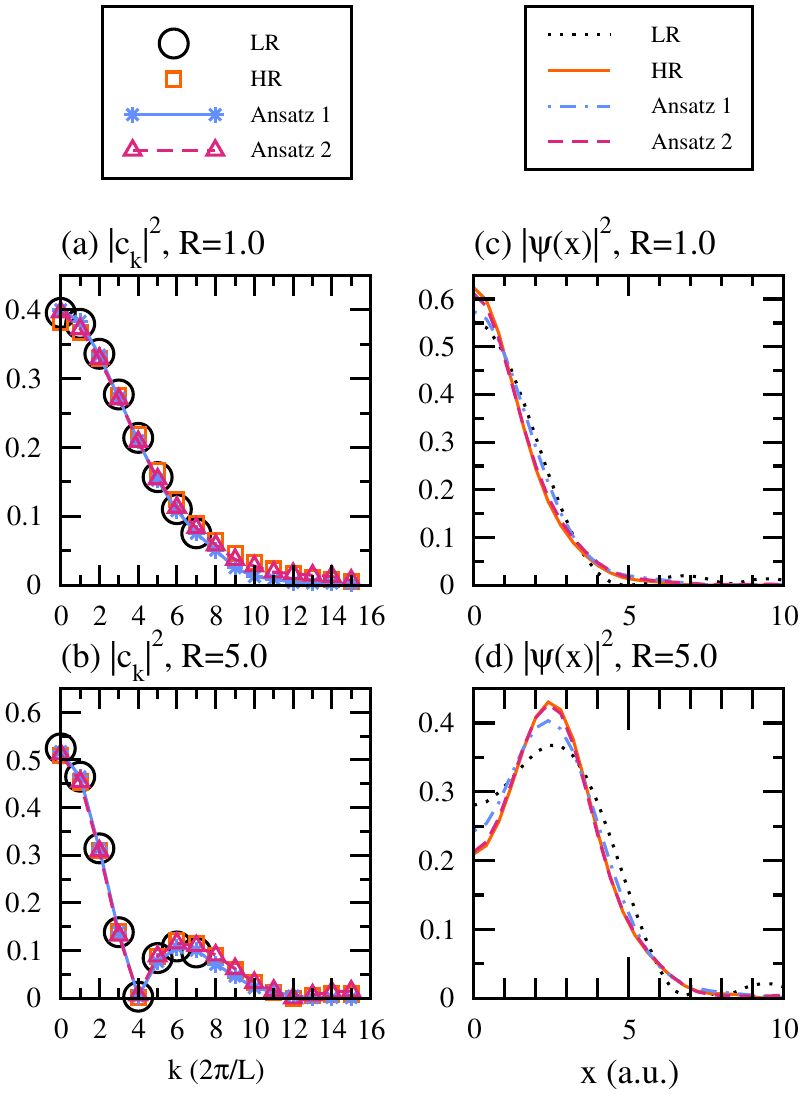}
\caption{The lowest occupied Hartree-Fock wavefunctions of the triplet
H$_{2}$ at bond length $R=1.0$ a.u.~(a, c) and $R=5.0$ a.u.~(b, d) in Fourier space (a, b) and in real space (c, d)
with $L=40$ a.u.~for $N_{\textrm{pw}}:15 \to 31$ resolution enhancement.}
\label{fig:wfn_h2_l40_15-31_triplet_1}
\end{figure}
\begin{figure}[tbp]
\centering
\includegraphics[clip]{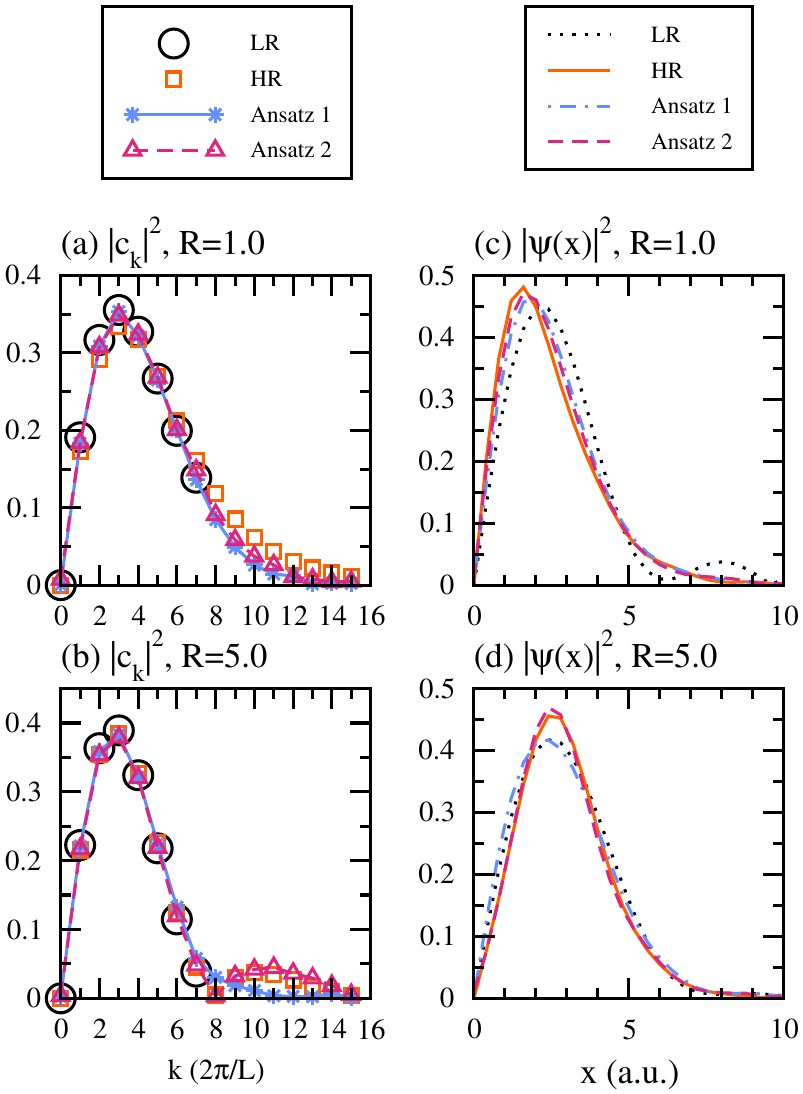}
\caption{The second lowest occupied Hartree-Fock wavefunctions of the triplet
H$_{2}$ at bond length $R=1.0$ a.u.~(a, c) and $R=5.0$ a.u.~(b, d) in Fourier space (a, b) and in real space (c, d)
with $L=40$ a.u.~for $N_{\textrm{pw}}:15 \to 31$ resolution enhancement.}
\label{fig:wfn_h2_l40_15-31_triplet_2}
\end{figure}
Figures \ref{fig:wfn_h2_l40_15-31_singlet}--\ref{fig:wfn_h2_l40_15-31_triplet_2}
show the results of the H$_{2}$ molecule
in case (ii).
In this case the LR Fourier wavevectors cover a wider region ($k\le 7$) and the contribution from
the tail region ($k > 7$) is small,
but one can still see that ansatz 2 shows a better agreement with the HR wavefunctions.

\begin{figure}[tbp]
\centering
\includegraphics[clip]{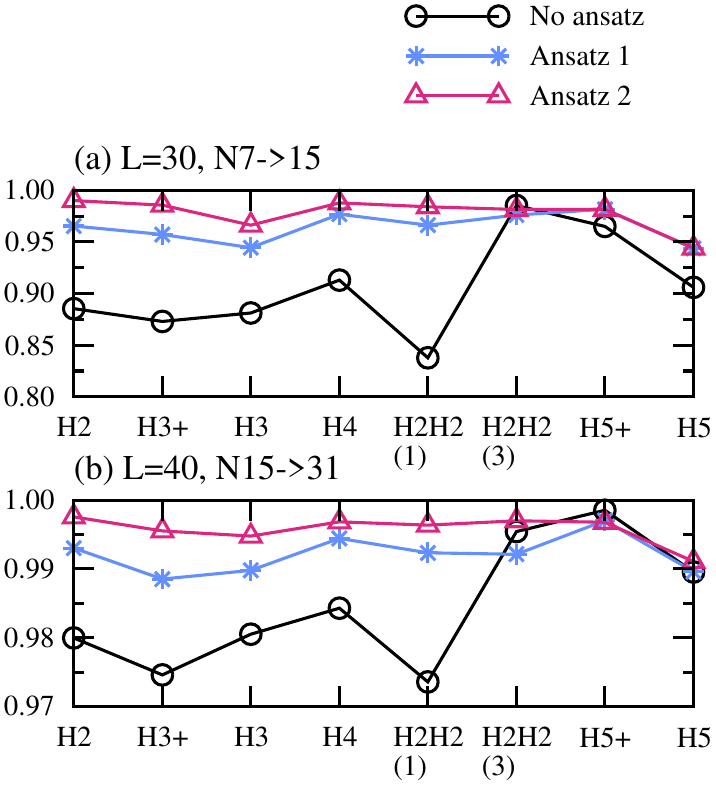}
\caption{Calculated average fidelity values for the validation dataset. The numbers in the parentheses of
the H$_{2}$H$_{2}$ labels indicate the bond length of the two H$_{2}$ molecules.}
\label{fig:fid_val}
\end{figure}

Figure \ref{fig:fid_val} shows the average fidelity values for each species included in the validation data
(Table \ref{tbl:valid}).
The validation dataset
contains samples for molecules and cations not included in the training data, namely H$_{3}^{+}$,
H$_{2}$-H$_{2}$(two H$_{2}$ molecules), H$_{5}^{+}$, and H$_{5}$.
One can see the same overall trend as the training data, where both ansatzes improve the fidelity, and
ansatz 2 generally gives higher fidelity values. Larger disagreements are seen in H$_{3}$ for case (i) and H$_{5}$,
both of which have odd numbers of electrons.
\begin{figure*}[tbp]
\centering
\includegraphics[clip]{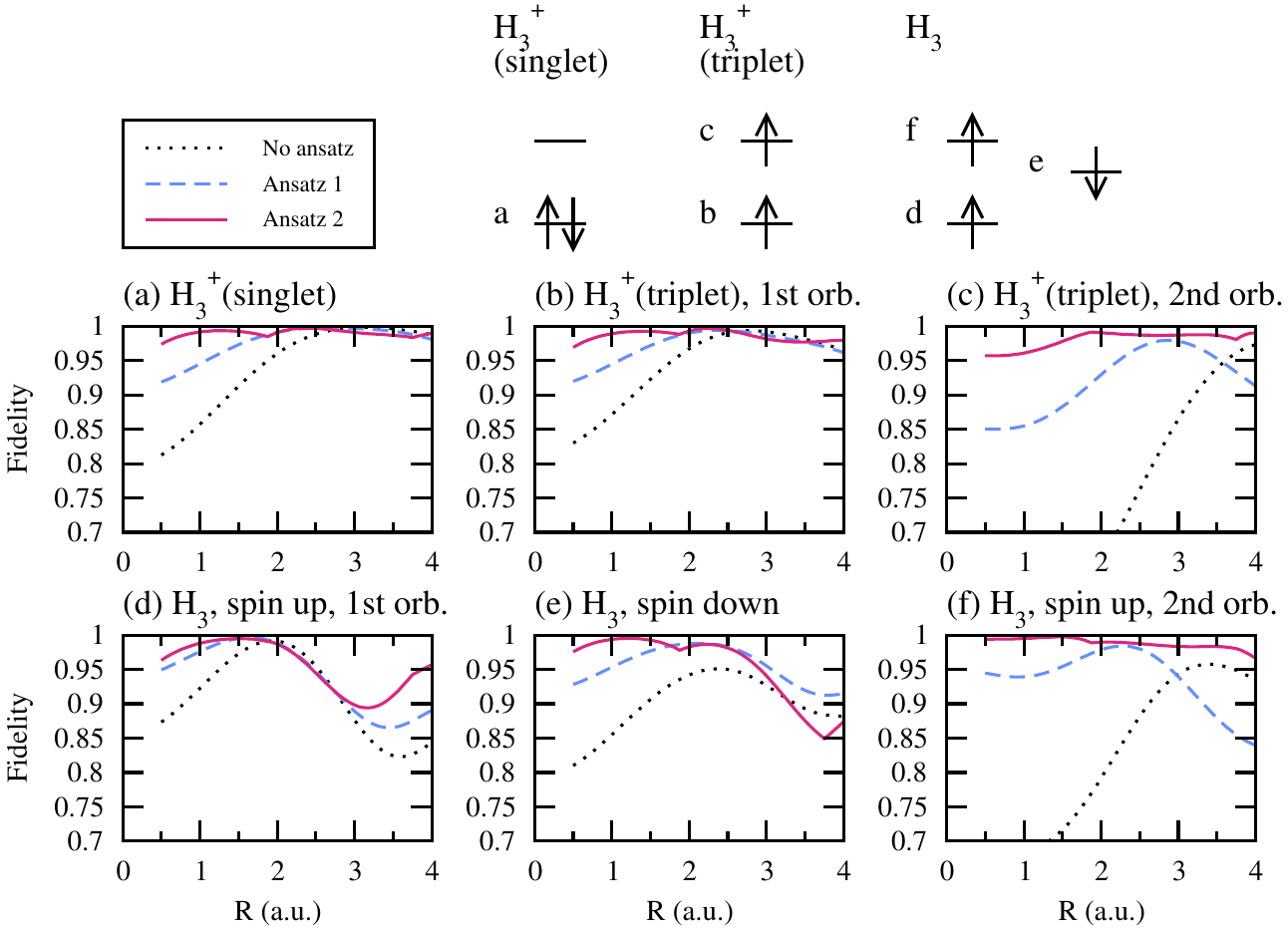}
\caption{Fidelity values of the occupied wavefunctions of H$^{+}_{3}$(a--c) and H$_{3}$(d--f) for various bond lengths
    for $N_{\textrm{pw}}: 7 \to 15$ resolution enhancement at $L=30$.}
\label{fig:fidelity_h3}
\end{figure*}
\begin{figure}[tbp]
\centering
\includegraphics[clip]{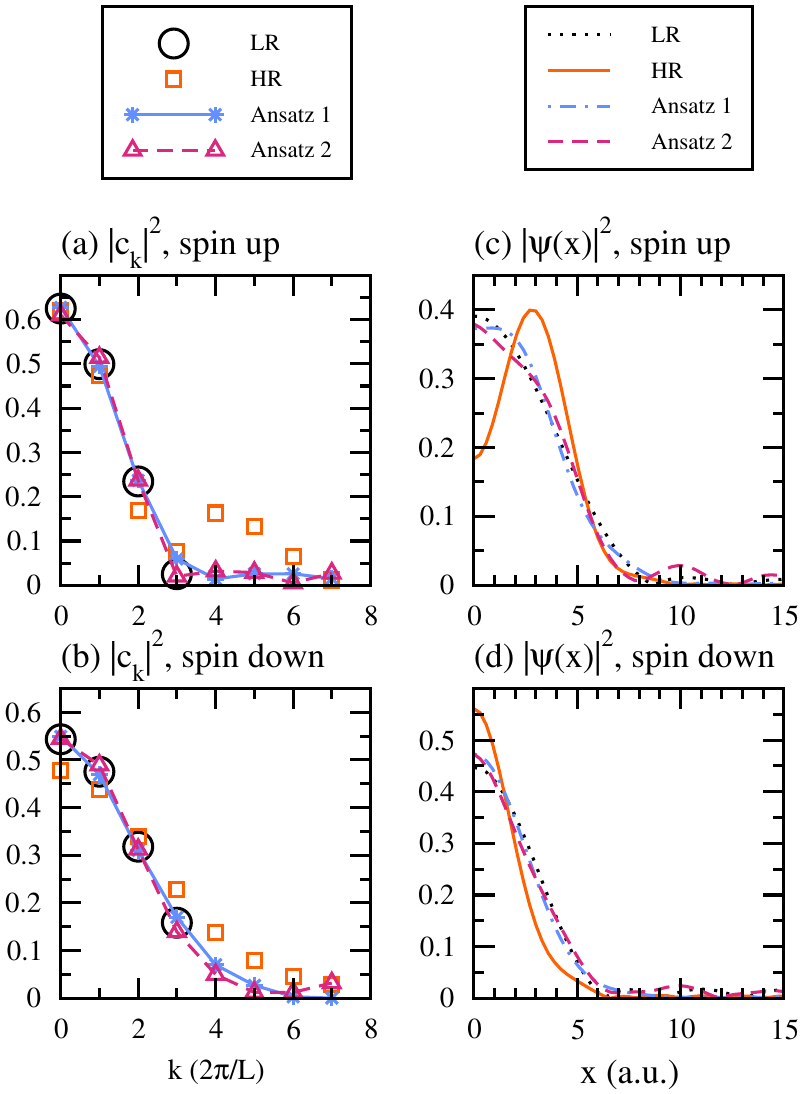}
\caption{The two lowest Hartree-Fock wavefunctions of H$_{3}$ at bond length $R=3.0$ at $L=30$
for $N_{\textrm{pw}}: 7 \to 15$ resolution enhancement. (a) the lowest occupied spin-up (majority spin) wavefunction,
    (b) the occupied spin-down (minority spin) wavefunction.}
\label{fig:wfn_h3}
\end{figure}
To analyze the obtained results, in Fig.~\ref{fig:fidelity_h3} we plot orbital-dependent
fidelity values of H$^{+}_{3}$ and H$_{3}$ for case (i) as a function of bond length $R$.
One can see that the LR wavefunctions
without including an ansatz, shown with dashed lines,
generally have low fidelity values for small $R$, due to the fact that in the LR calculation
the atomic positions cannot be properly resolved in this region. In fact, small $R$ seems to be the main domain where
the current QML models are the most effective.
Ansatz 2 yields higher fidelity values for the three H$^{+}_{3}$ orbitals (Fig.~\ref{fig:fidelity_h3} (a--c))
and the highest orbital of H$_{3}$ (Fig.~\ref{fig:fidelity_h3} (f)) for all $R$,
but large discrepancies are seen
in the lowest two orbitals of H$_{3}$ (Fig.~\ref{fig:fidelity_h3} (d, e)) for large $R$.
In Fig.~\ref{fig:wfn_h3} we plot these two wavefunctions at $R=3.0$.
While the HR curves of the two orbitals in the Fourier space (squares in Fig.~\ref{fig:wfn_h3} (a, b)) have qualitatively
different shapes especially for $k > 3$, the LR curves (open circles), whose domain is $k \leq 3$, are fairly similar,
making the prediction of distinct extrapolations rather challenging.
One possible way to improve the results is to include the spin dependence
in the effective potential used in Eq.~(\ref{eq:ansatz2}) for ansatz 2.
This point is left for future research.

H$_{5}$ is the largest molecule in the current dataset and is also the system where our models perform the poorest.
The reason for the low fidelity values may partly be due to the spin-dependence not included in our ansatz as
in the H$_{3}$ case,
and also because of its electron configuration. H$_{5}$ contains five occupied (spin) orbitals, with the highest one
having a distinctive spatial character that is not well represented in the training dataset. Adding training data with
this configuration would be required to improve the prediction performance on H$_{5}$.
Interestingly, the fidelity is higher for H$_{5}^{+}$ and also for two H$_{2}$ (H$_{2}$H$_{2}$), which
have the same electron configuration as H$_{4}$ and has no spin polarization.
These results suggest that the accuracy is more affected by
the electronic configuration of the system than by the atomic positions.

Our results indicate that ansatz 2 shows a good predictive power not only for the systems in the training dataset, but
also for those which were not present in the dataset, except when the electron configuration is significantly underrepresented in the
training dataset.
It may be interesting to interpret this ansatz in the context of quantum optimal control (QOC)
theory~\cite{Werschnik2007, Glaser2015, Mahesh2022};
QOC theory considers a unitary time propagator
with an effective Hamiltonian $e^{-i H_{\textrm{eff}} t}$,
where $H_{\textrm{eff}}$ consists of
the system Hamiltonian and the time-dependent drift or control Hamiltonian.
The latter Hamiltonian steers a given system
to some desired state, which in the current case is the true HR wavefunction.
Our ansatz 2, given by Eq.~(\ref{eq:ansatz2}), may be regarded as a QOC time propagator with
$N_{l}$ discretized time steps, where the potential term of the system Hamiltonian is explicitly included,
and the second part $(U_{\textrm{HEA}})$
takes into account the rest of the system Hamiltonian and also the (system-dependent) drift term.
Adding more layers in the ansatz corresponds to increasing the evolution time or
reducing the time step in the propagation, which is expected to give improved results.
One may be able to reduce the number of parameters in the ansatz by explicitly including the kinetic energy term.
It could also be possible to simplify the training algorithm
with the help of quantum optimal control techniques, such as the gradient ascent pulse
engineering (GRAPE) algorithm~\cite{Khaneja2005}.

\subsection{Two-electron case}

\begin{figure}[tbp]
\centering
\vspace{5mm}
\includegraphics[clip]{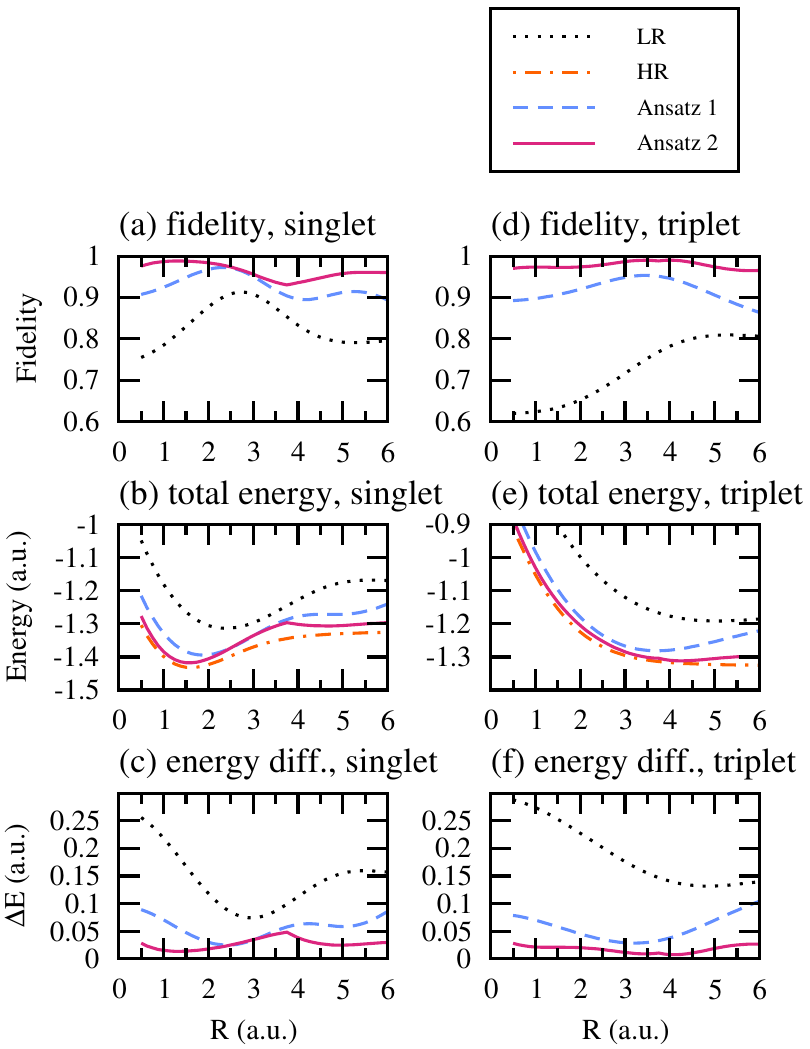}
\caption{Results of $N_{\textrm{pw}}: 7 \to 15$ enhancement
for the two-electron wavefunctions of the H$_{2}$ molecule at $L=30$ a.u.~as
a function of the bond length $R$ for the singlet state (a, b, c)
and the triplet state (d, e, f). (a) and (d) show the fidelity values between true $HR$ and
predicted wavefunctions, (b) and (e) show the total energies, and (c, f) are energy differences from HR results,
    $\Delta E = |E - E_{\textrm{HR}}|$.
}
\label{fig:h2_1}
\end{figure}
\begin{figure}[tbp]
\centering
\includegraphics[clip]{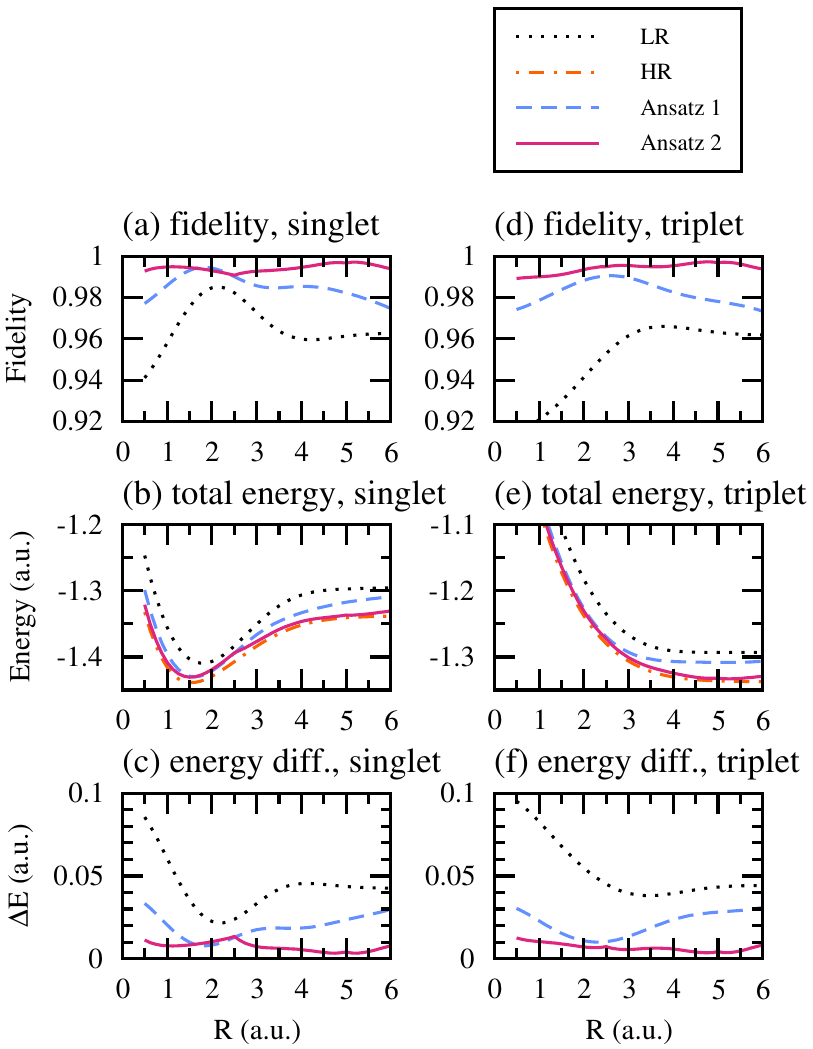}
\caption{Same as Fig.~\ref{fig:h2_1} but for $N_{\textrm{pw}}: 15 \to 31$ resolution enhancement at $L=40$.}
\label{fig:h2_2}
\end{figure}

To further investigate the performance of the trained models,
we apply the trained models to generate interacting HR two-electron wavefunctions of
the H$_{2}$ molecule from the LR wavefunctions,
using the approach explained in the previous section.
We consider the spatial part of two-electron wavefunctions in spin singlet ($s$) and
triplet ($t$) states
\begin{equation}
    |\Psi^{s, t}_{2}\rangle = \sum_{k_{0}k_{1}} C^{s, t}_{k_{0}k_{1}} |k_{0} k_{1}\rangle,
    \label{eq:twophi}
\end{equation}
where the expansion coefficients satisfy the following symmetry properties:
\begin{eqnarray}
    C^{s}_{k_{0}k_{1}} &=& +C^{s}_{k_{1}k_{0}}, \\
    \label{eq:singlet}
    C^{t}_{k_{0}k_{1}} &=& -C^{t}_{k_{1}k_{0}}.
    \label{eq:triplet}
\end{eqnarray}
In Figs.~\ref{fig:h2_1} (a, d) and Figs.~\ref{fig:h2_2} (a, d), the calculated fidelities are shown for various bond lengths.
It can be seen that as in the one-particle case, both of the trained models (ansatzes 1 and 2)
generate wavefunctions with
higher fidelity values with the true HR wavefunctions, and
ansatz 2 again yields better results.
Near the equilibrium bond length ($R\approx 1.5$ a.u.),
the wavefunction can be approximated by a single Slater determinant,
and by neglecting the antisymmetry of the wavefunctions
the fidelity values of the two-body wavefunctions $f$ can be
estimated from those of the one-particle orbitals in the wavefunction $f_{\mu\sigma}$
as $f \approx \prod_{\mu\sigma}^{\textrm{occupied}} f_{\mu\sigma}$. For case (i) (Fig.~\ref{fig:h2_1}),
the estimated fidelity values of
the singlet (triplet) H$_{2}$ at $R=1.0$ are 0.925 (0.900) and 0.991 (0.971) for ansatz 1 and 2, respectively,
and for case (ii) (Fig.~\ref{fig:h2_2})
they are 0.987 (0.980) and 0.997 (0.990) for ansatz 1 and 2, respectively.
The fidelity in ansatz 1 decreases for larger $R$ in both cases, which
again indicates the limitation of a linear ansatz for this problem. This error originates from
the inaccuracy of the one-particle model $U(\boldsymbol\theta)$ for larger $R$
as shown in, for example, one-particle data in Fig.~\ref{fig:wfn_h2_l30_07-15_triplet_1} (b).

In Fig.~\ref{fig:h2_1} and Fig.~\ref{fig:h2_2} we also compute the total energy of H$_{2}$,
calculated as the expectation value of
the following first-quantized Hamiltonian~\cite{Su2021}
\begin{eqnarray}
H &=& \sum_{i_{\textrm{el}}}^{2} \sum_{k, k'}\Bigl [
\delta_{k k'}\frac{k^{2}}{2} |k\rangle \langle k|_{i_{\textrm{el}}} \nonumber\\
&& -
\sum_{I}^{N_{\textrm{nuc}}}
Z_{I} e^{-i (k-k') d_{I}} \tilde{v}_{\textrm{exp}}(k-k')|k\rangle \langle k'|_{i_{\textrm{el}}}
    \Bigr ]  \nonumber\\
&& +   \frac{1}{2} \sum_{i_{\textrm{el}} \neq i_{\textrm{el}}'}^{2} \sum_{q, k, k'}
\Bigl [
\tilde{v}_{\textrm{exp}}(q) |k-q\rangle \langle k|_{i_{\textrm{el}}} |k'+q\rangle \langle k'|_{i_{\textrm{el}}'}
\Bigr ] \nonumber\\
&&    + E_{\textrm{nn}},
\end{eqnarray}
where $i_{\textrm{el}}, i_{\textrm{el}}'$ are the electron indices and $|k\rangle \langle k|_{i_{\textrm{el}}}$
act on the $i_{\textrm{el}}$-th register. The remainder of the equation is explained in Appendix \ref{app:pw}.
Although we do not optimize wavefunctions by minimizing the energy expectation values,
the calculated total energies with the generated wavefunctions are closer to the HR values.

It is interesting that this approach, using the models trained with one-particle wavefunctions,
also works for many-body wavefunctions, especially for larger $R$, where the wavefunctions have
a multi-reference character (i.e.~wavefunctions are more strongly entangled)
and mean-field approaches like the Hartree-Fock approximation become invalid.
This suggests one potential future application of the current approach,
which is to generate approximate many-body wavefunctions with low computational cost
by using a model $U(\boldsymbol\theta, \mathbf{D}^{(i)})$ trained with one-particle wavefunctions,
which can be obtained efficiently with classical computers.

\section{conclusions}
We performed a numerical experiment of quantum machine learning
with quantum input data, where one-particle wavefunctions expanded in the plane-wave basis are generated
from those composed from a smaller number of basis functions.
The results were improved significantly by including
sample-dependent information in the ansatz, suggesting
the importance of nonlinearity in the ansatz for this problem.
The trained models yielded reasonable results also for molecular structures not
included in the training dataset, although they do not generalize enough to generate HR wavefunctions
with unseen electronic configurations. Possible pathways to improve the generalizability have been identified.
We also showed that the trained models can be used to enhance the resolution of two-electron wavefunctions,
and that many-body correction is required for a scalable generalization of the current approach
to many-body wavefunctions.

The current work provides one QML example where
a data re-uploading-like ansatz is successfully combined with quantum input,
by injecting data-dependent classical information into the ansatz.
It may be interesting to see if a similar improvement can be achieved in other cases,
when some auxiliary data-dependent information is available.

One may think of potential applications of this approach in various directions,
such as the generation of all-electron wavefunctions from pseudo wavefunctions
in pseudo-potential-based calculations~\cite{Martin2004}.

\appendix
\section{Plane-wave basis}
\label{app:pw}
We consider the following one-dimensional Hamiltonian for $N_{\textrm{el}}$ electrons in the presence of
$N_{\textrm{nuc}}$ nuclei
\begin{eqnarray}
H &=& \sum_{i=1}^{N_{\textrm{el}}}
\Bigl [ -\frac{1}{2}\frac{d^2}{dx_{i}^2} + v_{\textrm{en}}(x_{i}) \Bigr ] +
\sum_{i < j}^{N_{\textrm{el}}}  V_{\textrm{ee}}(x_{i} - x_{j}) \nonumber\\
&& + E_{\textrm{nn}},
\label{eq:hamiltonian}
\end{eqnarray}
where $v_{\textrm{en}}(x)$ is the attractive Coulomb potential between the electrons and the nuclei, 
$V_{\textrm{ee}}(|x_{i} - x_{j}|)$ is the repulsive Coulomb potential between the electrons,
and $E_{\textrm{nn}}$ is the
Coulomb potential between the nuclei, which is a constant under the Born-Oppenheimer approximation.
We use atomic units throughout.
The explicit forms of these terms are
\begin{equation}
    \label{eq:el-nuc}
    v_{\textrm{en}}(x) = \sum_{I}^{N_{\textrm{nuc}}} -Z_{I} v_{\textrm{exp}}(x - d_{I}),
\end{equation}
\begin{equation}
    V_{\textrm{ee}}(x) = v_{\textrm{exp}}(x),
\end{equation}
\begin{equation}
    E_{\textrm{nn}} = \sum_{I<J} Z_{I}Z_{J}v_{\textrm{exp}}(d_{I}-d_{J}).
\end{equation}
Here $Z_{I}$ and $d_{I}$ are respectively the atomic number and the position of nucleus $I$, and
$v_{\textrm{exp}}(x)$ is the exponential Coulomb-mimicking potential proposed in Ref.~\cite{Baker2015}
\begin{equation}
    v_{\textrm{exp}}(x) = A \exp(-\kappa |x|)
\end{equation}
with $A=1.071295$ and $\kappa=1 / 2.385345$.

The Hartree-Fock one-particle Hamiltonian in Eq.~(\ref{eq:hf}) for spin $\sigma$$ (=\uparrow, \downarrow )$
is given as
\begin{eqnarray}
    h^{\textrm{HF}\sigma}_{jj'} &=& \langle \chi_{k_{j}}|h^{\textrm{HF}\sigma}|\chi_{k_{j'}}\rangle \nonumber\\
    &=& h^{0}_{jj'} + \sum_{\sigma'}\sum_{ll'}
    D_{ll'}^{\sigma'}\Bigl [ V_{jj'l'l} - \delta_{\sigma\sigma'}V_{jll'j'} \Bigr ],
\end{eqnarray}
where
\begin{eqnarray}
    h^{0}_{jj'} &=& \delta_{k_{j}k_{j'}} \frac{k_{j}^{2}}{2} \nonumber\\
    && - \sum_{I}^{N_{\textrm{nuc}}}Z_{I}e^{i(k_{j}-k_{j'})d_{I}}
    \tilde{v}_{\textrm{exp}}(k_{j}-k_{j'}),
\end{eqnarray}
\begin{equation}
    D_{ll'}^{\sigma} = \sum_{\mu}^{\textrm{occupied}} C_{k_{l}\mu\sigma} C^{*}_{k_{l'}\mu\sigma},
\end{equation}
\begin{equation}
V_{jj'll'} = \delta_{k_{j}+k_{j'}, k_{l}+k_{l'}} \tilde{v}_{\textrm{exp}}(k_{l} - k_{l'}).
\end{equation}
Here $\tilde{v}_{\textrm{exp}}(k)$ are the Fourier components of $v_{\textrm{exp}}(x)$
\begin{equation}
    \tilde{v}_{\textrm{exp}}(k) = \frac{2 A \kappa}{L(\kappa^{2} + k^{2})}.
\end{equation}

\section{Dataset}
\label{app:dataset}
We prepare one-particle wavefunctions of $H_{x}$
molecules ($x=1, 2, 3, 4, 5$) and cations in the spin-restricted Hartree-Fock
(RHF) or unrestricted Hartree-Fock (UHF) approximations,
which are summarized in Tables \ref{tbl:train} and \ref{tbl:valid} for
the training and the validation datasets, respectively.

\begin{table}[htb]
\begin{ruledtabular}
\begin{tabular}{l c c c}
& method & $N_{\textrm{orb}}$ & $R$ (a.u.) \\ \hline
H$_{2}$ (singlet) & RHF & 1 & [1.0, 2.0, 3.0, 4.0, 5.0, 6.0] \\
H$_{2}$ (triplet) & UHF & 2 & [1.0, 2.0, 3.0, 4.0, 5.0, 6.0] \\
H$_{3}$ & UHF & 3 & [1.0, 2.0, 3.0, 4.0] \\
H$_{4}$ & RHF & 2 & [1.0, 2.0, 3.0, 4.0] \\
\end{tabular}
\caption{Training dataset used in this work. Here $N_{\textrm{orb}}$ is the number of orbitals used
and $R$ is a bond length. All H atoms are equidistantly aligned.}
\label{tbl:train}
\end{ruledtabular}
\end{table}
\begin{table}[htb]
\begin{ruledtabular}
\begin{tabular}{l c c c}
& method & $N_{\textrm{orb}}$ & $R$ (a.u.) \\ \hline
H$_{2}$ (singlet) & RHF & 1 & [0.5, 0.5625, 0.625, $\cdots$, 6.0] \\
H$_{2}$ (triplet) & UHF & 2 & [0.5, 0.5625, 0.625, $\cdots$, 6.0] \\
H$^{+}_{3}$ (singlet) & RHF & 1 & [0.5, 0.5625, 0.625, $\cdots$, 4.0] \\
H$^{+}_{3}$ (triplet) & UHF & 2 & [0.5, 0.5625, 0.625, $\cdots$, 4.0] \\
H$_{3}$ & UHF & 3 & [0.5, 0.5625, 0.625, $\cdots$, 4.0] \\
H$_{4}$ & RHF & 2 & [0.5, 0.5625, 0.625, $\cdots$, 4.0] \\
H$_{2}$-H$_{2}$ & RHF & 2 & [0.5, 0.5625, 0.625, $\cdots$, 4.0] \\
H$_{5}^{+}$ & RHF & 2 & [1.5, 1.5625, 1.625, $\cdots$, 3.0] \\
H$_{5}$ & UHF & 5 & [1.5, 1.5625, 1.625, $\cdots$, 3.0] \\
\end{tabular}
\caption{Validation dataset used in this work. All H atoms are aligned equidistantly except in H$_{2}$-H$_{2}$, where
$R$ indicates the distance between two H$_{2}$ molecules with bond length $R'$.}
\label{tbl:valid}
\end{ruledtabular}
\end{table}

We preprocess the phase of the training dataset so that
$C_{k_{0}\mu\sigma} > 0$ is satisfied for all the LR and HR wavefunction coefficients,
where $k_{0} = \frac{2 \pi}{L}$ is an arbitrarily chosen
wavevector. For the molecules in the training data with $N_{\textrm{orb}}\geq 2$
(H$_{2}$(triplet), H$_{3}$ and H$_{4}$),
we additionally include $\psi_{\mu_{12}\sigma} = \frac{1}{\sqrt{2}}\bigl [ \psi_{\mu_{1}\sigma} + \psi_{\mu_{2}\sigma} \bigr ]$ in the training
datasets to align the phases of the predicted HR wavefunctions, as explained in the main text.

\section{Linear interpolation}
\label{app:intp}
\begin{figure*}[tbp]
\centering
\includegraphics[clip]{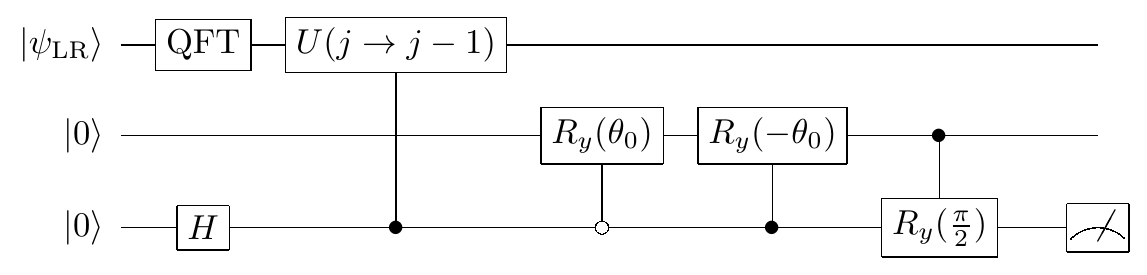}
\caption{Quantum circuit for the linear interpolation. The states with $0$ in the ancilla qubit (bottom) are post-selected.
Here $\theta_{0}=2\arctan(\frac{1}{\sqrt{2}})$.}
\label{fig:qc_intp}
\end{figure*}
We consider a linear interpolation of a one-dimensional vector $f \in \mathbb{C}^{N}$ encoded
in the amplitudes of $n=\log_{2} N$ qubit state as
$f=\sum_{j=0}^{N-1}f_{j}|j\rangle = \sum_{j_{0},j_{1},\dots,j_{n-1}} f_{j_{0}j_{1}\dots j_{n-1}}|j_{0}\rangle \otimes \dots \otimes |j_{n-1}\rangle$,
where $j=j_{n-1} + 2 j_{n-2} + \cdots + 2^{n-2} j_{1} + 2^{n - 1} j_{0}$.
Our objective is to generate $2N$ states with $n+1$ qubits as
\begin{eqnarray}
    && \sum_{j=0}^{N - 1} f_{j}|j\rangle \otimes |0\rangle \nonumber\\
    && \to \mathcal{N}
    \sum_{j=0}^{N-1} \Bigl [ f_{j}|j\rangle |0\rangle + \frac{f_{j} + f_{j+1}}{2}|j\rangle |1\rangle
    \Bigr ],
    \label{eq:intp}
\end{eqnarray}
where $\mathcal{N}$ is a normalization factor, and we define $f_{N} = f_{0}$ (periodic boundary condition).
We perform the linear interpolation of the one-particle wavefunctions $\{\psi_{\mu\sigma}\}$
in real space with one ancilla qubit, using the quantum circuit shown in Fig.~\ref{fig:qc_intp}.
The first QFT transforms the wavefunctions to real space, and $U(j\to j-1)$ is an operator that shifts
$|j\rangle$ to $|j-1\rangle$ ($|-1\rangle = |N - 1\rangle$), which can be implemented as described in
Ref.~\cite{Fan2016}.
It should be noted that this interpolation breaks the orthonormality of the interpolated wavefunctions.

\end{document}